\documentclass[12pt,prl]{revtex4}

\usepackage{amsmath}

\newcommand{\cam}{{\cal M}}
\newcommand{\ba}{{\bf A}}
\newcommand{\bb}{{\bf B}}

\begin{document}
\title{Observed Quark Mixing Structure (CKM-Matrix) -- Perception of
Mirror Generations}

\author{I. T. Dyatlov\\
\em Petersburg Nuclear Physics Institute,\\
RNC ``Kurchatov Institute",
St.~Petersburg, 188300 Gatchina, Russia }
\date{}

\begin{abstract}
Parity nonconservation led Lee and Yang (1956) to the hypothesis of
additional fermions with properties that mirror those of usual
particles. It is shown here that the well-known observed qualitative
structure of the quark mixing matrix (CKM matrix) is precisely
reproduced when usual quark mass hierarchies result from mass
hierarchies of heavy mirror generations. The last ones are of the
Lee--Yang type only. The usual quark spectrum is formed likewise the
see--saw spectrum of the neutrino physics. The lightest mirror quark is
closely related with the $t$ quark and might be available even in a
region below 1~TeV.

\end{abstract}
\maketitle


The interrelation between quark masses and the weak mixing matrix
\cite{1} has already been considered over several decades. The case is
not a simple representation for both matrices by means of the same
parameters. In general, such an operation could always be realized with
potentially 36 independent variables of two mass
matrices for up $(\bar u)$ and down $(\bar d)$ quarks. Parameter choice
and expressions for observables should, as such, immediately exhibit
relations between mass and mixing hierarchies and, at best, help to
clarify mechanisms of quark mass formation.

An active search for interrelation of structures was initiated by
Fritzsch in 1977--1978 \cite{2}. An important step in the problem
was his suggestion to use the so-called ``democratic matrix" (a matrix
with all equal elements) when considering hierarchies \cite{3}. After
that these matrices appeared in a large number of papers concerning the
problem \cite{4}. Proposed in \cite{5,6} parametrizations (adopting in
their forms observed hierarchies of CKM elements \cite{1}) were also
exceptionally useful, especially the Wolfenstein one \cite{6}.

In this work, the mass-matrix representation (or parametrization
alternative) is suggested that, without a numerical choice for
elements, leads to the well-known observed structure of the CKM matrix
and to the unambiguous physical interpretation both for mass and mixing
forms.

The suggested parametrization  involves a sum of the matrices
$\cam^{(n)}_{LR}$, $n=0,1,2$, each with factorized matrix elements. The
entire mass matrix is given by the formula $(\cam=\cam_{LR}+\cam_{RL})$
\begin{equation}
(\cam_{LR})^b_a\ =\ \sum_{n=0,1,2}(\cam^{(n)}_{LR})^b_a\ =
\sum_{n=0,1,2} A^n_a B^{+b}_n\,.
\end{equation}
Here $a$ and $b$ are indices of quark generations. Vectors in $a,b$
space: $\ba$ and $\bb$ -- have arbitrary complex components. In what
follows vector notations will be omitted: $|\ba|\equiv|A|=(A^+A)^{1/2}$
and so on.

The overall number of parameters in Eq.\,(1) large as it is,
nevertheless is not so large as one could expect calculating vector
components.  Physical quantities do not depend on $a,b$ indices, some
phases can be taken up by quark operators. But the main point, however,
is that all results will not depend on numbers and ambiguous parameter
choices.  They are due exclusively to the very form (1), its separable
structure. The real parameter number may come out from physical
requirements (see further).

Every matrix $\cam^{(n)}_{LR}$ is a direct generalization of the
``democratic" matrix: it also has only one nonzero eigenvalue. All
three eigenvalues become unequal to zero in the separable form (1). The
hierarchical steps for quark masses are introduced by the relations
between factors $A$ and $B$ with distinct $n$
\begin{equation}
|A_0|,|B_0|\ \gg\ |A_1|,|B_1|\ \gg\ |A_2|,|B_2|.
\end{equation}
Eqs. (2) are preliminary assumptions, our intention is to relate them
with the known structure of the CKM matrix.

The relations (2) permit to obtain widely separated quark masses
\begin{equation}
m_I\ \gg\ m_{II}\ \gg\ m_{III}
\end{equation}
after diagonalization of the matrix (1).
As usual, one diagonalizes the hermitian matrix $(\cam\cam^+)_{LL}$ and
obtains eigenvalues $m^2_T$.

The $L$ eigenfunctions for $(\cam\cam^+)^{(f)}_{LL}$ with different
flavours $f=\bar u,\bar d$ are necessary to construct the CKM matrix,
corresponding to Eq.\,(1)
\begin{equation}
C_{TS}\ =\ \Big(\Phi^{(\bar u)+}_{LT}, \Phi^{(\bar d)}_{LS}\Big).
\end{equation}
In this formula, we have $T,S=I,II,III$. Numbers correspond to
decreasing masses. $I$ is the heaviest generation, i.e. the third one
in the standard numeration $(t,b)$.

The matrix (4) has no remarkable properties when $A,B$ quantities are
represented by arbitrary values (besides (2)). Eq.~(4) presents then
practically arbitrary unitary matrix.

The situation is drastically changed when one takes the vector $A$ as
independent of $\bar u,\bar d$ indices. So, if the equality
\begin{equation}
A^{(\bar u)}\ =\ A^{(\bar d)}
\end{equation}
is imposed, one can consider the factor $A$ as invariant under the weak
$SU_L(2)$ symmetry. The matrix (4) immediately acquires all qualitative
properties of the CKM \cite{1} and Wolfenstein \cite{6,1} matrices. The
relation
\begin{equation}
C^+_{ts}C_{cd}\ =\ C_{td}+C^+_{ub}
\end{equation}
holds. This relation is the direct result of all properties inherent in
the approximate Wolfenstein matrix (see \cite{1}): the hierarchy
orders, equalities:
\begin{equation}
C_{ud}\approx C_{cs}\approx C_{tb}\approx1\,, \quad C_{cd}=-C_{us}\,,
\quad C_{ts}=-C_{cb}\,,
\end{equation}
and orthogonality of its first and third columns (rows).

No additional conditions are required to fulfill Eqs. (6) and (7). No
preliminary choice for $A$ and $B$ components is necessary (naturally,
besides Eqs. (2) and (5)). One obtains (6) and (7) automatically, for
any values of $A$ and $B$ vectors and with their arbitrary phases.

General formulae for the standard diagonalization procedure with the
matrix (1) are too cumbersome. When taking only the lowest hierarchy
approximation, only one mass becomes finite. Next orders need to be
calculated to split two other degenerate levels. Besides, the second
order contributions to $C_{I\,III}$ (i.e. $C_{td}$ or $C_{ub}$) cancel;
it is necessary to find and use second and third order terms in masses
and eigenfunctions. These are the sources of complications. Only
relevant terms of hierarchy orders are written out; unimportant parts
of the same order are omitted. The extended version with all
contributions will be published elsewhere.

Masses of the $I-III$ states are
\begin{equation}
m^2_I\,=|A_0|^2|B_0|^2+2\mbox{Re}\Big[(A^+_0,A_1)(B_0,B^+_1)\Big]
+2\mbox{Re}\Big[(A^+_0,A_2)(B_0,B^+_2)\Big]+\cdots,
\end{equation}

\begin{eqnarray}  
m^2_{II} &=& \frac{D_2(|A_0|^2,|A_1|^2)D_2(|B_0|^2,|B_1|^2)}{|A_0|^2|B_0|^2}
\bigg\{ 1- \frac{2\mbox{Re}[(A^+_0,A_1)(B^+_1,B_0)]}{|A_0|^2|B_0|^2}\ +
\nonumber\\
&&+\ \frac{2\mbox{Re}[D_2(|A_0|^2,(A^+_1,A_2))D_2(|B_0|^2,(B^+_2,A_1))]}{
D_2(|A_0|^2,|A_1|^2)D_2(|B_0|^2,|B_1|^2)} + ... \bigg\} ;
\end{eqnarray}

\begin{eqnarray}  
m^2_{III} &=& \frac{D_3(|A_0|^2,|A_1|^2,|A_2|^2)D_3(|B_0|^2,|_1|^2,|B_2|^2)}{
D_2(|A_0|^2,|A_1|^2)\,D_2(|B_0|^2,|B_1|^2)}\ \times
\nonumber\\
&&\times\ \bigg\{1- \frac{2\mbox{Re}[D_2(|A_0|^2,(A^+_1,A_2))
D_2(|B_0|^2,(B^+_2,B_1))]}{ D_2(|A_0|^2,|A_1|^2)D_2(|B_0|^2,|B_1|^2)}
+...\bigg\}.
\end{eqnarray}
In these formulae factors $D_2$ and $D_3$ are the determinants of the
second and third orders, constructed out of scalar products for
$A_n,A^+_n,B_n,B^+_n$ vectors. One has the following formula as $D_2$
\begin{equation}
D_2\Big((a,b),(c,d)\Big)\ =\ \left| \begin{array}{cc}
(a,b) & (c,b)\\
(a,d) & (c,d) \end{array} \right|.
\end{equation}
Only the diagonal elements are preserved in the arguments of
determinants. $D_3$ is defined similarly using scalar products of six
vectors.

The first term in Eq. (8) is the lowest approximation for heaviest mass
(the first $(I)$ generation in our notation). The factors before figure
brackets are the same for the $II$ and $III$ generations.

Orthonormal wave functions of eigenstates are found using the expansions
(8)--(10) and the matrix elements of the $(\cam\cam^+)_{LL}$ matrix.
These functions are vectors in the space of generation indices. The
following equations represent the approximations for $L$ functions that
are necessary for verifications of Eqs. (6) and (7):

\begin{equation}
\Phi_I\ =\ \frac1{|A_0|}\left\{ A_0+\frac{(B^+_1,B_0)}{|B_0|^2} \Big(1+
\frac{(2\mbox{Re}((A_0,A^+_1)(B^+_0,B_1))}{|A_0|^2|B_0|^2}\Big)A_1
+\frac{(B^+_2,B_0)}{|B_0|^2}A_2 \right\} ;
\end{equation}
\begin{eqnarray}
\Phi_{II} &=& \frac{|A_0|}{\{D_2(|A_0|^2,|A_1|^2)\}^{\frac12}} \Bigg\{
A_1-\frac{(A^+_0,A_1)}{|A_0|^2}A_0 -\frac{D_2(|A_0|^2,|A_1|^2)}{|A_0|^4}
\frac{(B_0^+,B_1)}{|B_0|^2}A_0 -
\nonumber\\
&&-\ \frac{D_2(|B_0|^2,(B^+_2,B_1))}{D_2(|B_0|^2,|B_1|^2)}
\frac{[A^+_0[A_0,A_2]]}{|A_0|^2} \Bigg\};
\end{eqnarray}

\begin{eqnarray}
\Phi_{III} &=&  \frac1{\{D_2(|A_0|^2,|A_1|^2)\}^{\frac12}} \Bigg\{
[A^+_0,A_1^+]+ \frac{D_2(|B_0|^2,(B^+_1,B_2))}{D_2(|B_0|^2,|B_1|^2)}
[A^+_0,A^+_2]\ +
\nonumber \\
&& +\  \frac{D_2((B^+_0,B_1),(B^+_1,B_2))}{D_2(|B_0|^2,|B_1|^2)} \,
[A^+_1,A^+_2] \Bigg\}.
\end{eqnarray}

Square brackets imply vector products. Normalization factors are
written out only in the lowest approximations, which is sufficient for
our purpose. All diagonal elements of the $C_{TS}$ matrix (4) are
approximately equal to one when, in the lowest order, the wave
functions (12)--(14) become independent of flavour indices, i.e. at
$A^{\bar u}_n\equiv A^{\bar d}_n$. Simultaneously, mutual orthogonality
$\Phi^{\bar u}_T$ and $\Phi^{\bar d}_S$, $T\neq S$, in the same order
makes evident a smaller size of nondiagonal elements.

Hence, the hierarchy of $C_{TS}$ elements arises and the condition (5)
is of a paramount importance in the process.

The mixing matrix elements can be calculated directly with Eqs.
(12)--(14). The sum $(C_{td}+C^+_{ub})$, in agreement with Eq.\,(6),
will give precisely the product $C_{ts}^+\cdot C_{cd}$ for any complex
$A$ and $B$ vectors satisfying Eq.~(5). We do not write out general
formulae because of their large sizes. However, for real vectors $A,B$
the interconnection of hierarchies can be illustrated in simple and
informative formulae. The elements $C_{ts}$ and $C_{cd}$ (real in the
Wolfenstein matrix, see \cite{1}) are well suited for that aim.

Real  vectors $A_n$ and $B_n$ are characterized by angles between them
(with different $n$): $\alpha_{in}$ and $\beta_{ik}$ corerspondingly,
-- and their lengths. Extracting quark masses (8)--(10) from equations
for CKM matrix elements, one obtains:
\begin{equation}
|C_{ts}|=\frac{m_s}{m_b}(\cot\beta_{01})_{\bar
d}-\frac{m_c}{m_t}(\cot\beta_{01})_{\bar u} \simeq
(0.025-0.02)(\cot\beta_{01})_{\bar d}-0.08(\cot\beta_{01})_{\bar u}\,,
\end{equation}
\begin{eqnarray}
&&|C_{cd}|\,=\,\frac{m_d}{m_s}f(\beta_{\bar d})
-\frac{m_u}{m_c}f(\beta_{\bar u})\,\simeq\,(0.06-0.05)
f(\beta_{\bar d})-0.0025 f(\beta_{\bar u}),
\\
&& f(\beta)=\frac{\cos\beta_{12}
-\cos\beta_{01}\cos\beta_{02}}{\sin\beta_{12}\,\cos\beta_{012}}\,,\
\sin^2\beta_{012}=\frac{\cos^2\beta_{01}+\cos^2\beta_{02}
-2\cos\beta_{12}\cos\beta_{01}\cos\beta_{02}}{\sin^2\beta_{12}}\,.
\nonumber
\end{eqnarray}
Experimental values for both $C$ are: $|C_{ts}|\approx0.04$,
$C_{cd}|\approx0.225$. Here $\beta_{012}$ is the angle between the
perpendicular to the plane $B_1B_2$ and the vector $B_0$. The product
Eq.~(15) $\times$ Eq.~(16) again gives the calculated sum
$C_{td}+C^+_{ub}$ for real vectors $A_n,B_n$.

The abundance of parameters makes numerical comparisons ineffective.
However, it is interesting to notice that $\beta_{01}\approx\pi/6$.

The elements $C_{ts}$ and $C_{cd}$ are both the first order quantities
of hierarchy steps:\mbox{ $\sim\!(m_{II}/m_I)$} and $(m_{III}/m_{II})$.
This property makes them different from the hierarchy proposed by
Wolfenstein. In the latter case one has in terms of the small parameter
$\lambda$: $|C_{cd}|\!\sim\!\lambda$, $|C_{ts}|\!\sim\!\lambda^2$. It
can be noticed that with the experimental values of masses the Cabibbo
element $C_{cd}$ will be larger than $C_{ts}$. But in our presentation
difference between them is only numerical. Quark masses depend on an
energy scale; such dependence does not significantly change the
ratios of masses.

The physical interpretation of Eq.~(1) is evident: as always,
factorization implies that transitions $R\to L$ go through one particle
states $n$. At the same time the generation indices can change:
$a\leftrightarrow b$. The quantum numbers (besides the generation ones)
of intermediate states $n$ should be the same as quark numbers.
Therefore these states are states of new quark generations with color,
spinor properties, flavor, and so on.

An energy scale attributed to the $n$ state will be named ``mass"
$M_n$, keeping in mind the simplest picture, as if the transition
proceeded through a free fermion propagator. The factorized form arises
when $M_n\gg m_T$; $m_T$ are masses of observed quarks. Then, in the
region $\hat p\simeq m_T$ the $n$ propagator becomes
\begin{equation}
\frac1{M_n-\hat p}\ \approx\ \frac1{M_n}\,.
\end{equation}
It is these masses $M_n$ that could determine values of the hierarchy
steps (2). The redefined vectors $(\tilde A,\tilde B)=(A,MB)$ may in
this case have the same magnitude for all $n$. Then, the quark masses
are ordered inversely in respect to the $M_n$ ones.

The components of $A_n$ and $B_n$ are now the conversion coefficients
for $a\to n$ and $n\to b$ transitions $(\psi\to\Psi)$. The mass-matrix
is
\begin{equation}
\cam^b_a\ =\ \sum_n \,\frac{\tilde A^n_a\,\tilde B^{+b}_n}{M_n}\,.
\end{equation}

In their turn, the relations (5) let one determine $SU_L(2)$ properties
of $n$ states. Transitions $a\to n$, $n\to b$ imply that the terms
\begin{equation}
\tilde A_a^{(f)n}\bar\psi_L^{(f)a}\Psi^{(f)}_{Rn}\,, \quad
\tilde B_n^{(f)+b}\bar\Psi_L^{(f)n}\psi^{(f)}_{Rb}
\end{equation}
and their complex conjugates are present within the Lagrangian. The
vector $\tilde A^{(f)}$ is independent of flavor $f$. Then, the first
term (19) becomes an invariant of $SU_L(2)$ when $R$ doublets
$\Psi^{(f)}_{Rn}$ are transformed in the same way with $L$ doublets of
usual quarks, i.e.
\begin{equation}
\Psi'_{Rn}\ =\ U_L(2)_n^{n'}\Psi_{Rn'}\,.
\end{equation}

Provided that all interactions, including the weak one, are symmetric
under simultaneous $R\leftrightarrow L$, $\psi\leftrightarrow \Psi$
transformations, the fermions $\Psi_n$ are called the ``mirror
particles" with respect to the usual ones $\psi_a$. The mirror
fermions, their role and possible existence have been discussed by many
authors, starting with the first papers by Lee and Yang
\cite{7} on parity violation. This discussion was continued in a
variety of directions: the mirror world interacting with our Universe
just gravitationally (see \cite{8} with huge bibliography), mirror
particles with large masses and varied weak properties (see review
\cite{9}), even to the point of their possible observation by LHC
\cite{10}.

Eq. (5) sets off the Lee--Yang variant from the whole diversity of the
``mirror ideas". With Lee and Yang one has common weak interactions for
usual and mirror particles, or, using the modern language, common $W$
bosons and the same group $SU_L(2)$. Such mirror generations could be
involved according to (5) and (20) in creation of $\cam^b_a$ and
$C_{TS}$ structures.

In a mirror-symmetrical system matrices $\tilde A$ and $\tilde B$ are
to be hermitian. So the expressions (19) can be rewritten in terms of
mirror-symmetrical operators:
\begin{equation}
\Psi^{(+)}_{RL\,na}=\Psi_{Rn}+\psi_{La}\,, \quad
\Psi^{(+)}_{LR\,na}=\Psi_{Ln}+\psi_{Ra}\,,
\end{equation}
and  can be diagonalized to give the Dirac type mass terms for states:
\begin{equation}
\Psi^{(1)}_n=(\Psi_R+\psi_L)_n\,, \quad
\Psi^{(2)}_n=(\Psi_L+\psi_R)_n\,.
\end{equation}
Thus, the separation into $A$ and $B$ in Eq.\,(1) acquires the physical
meaning. They can be associated with two mass-terms  in the
mirror-symmetrical Lagrangian. There are six massive states for every
$\bar u,\bar d$ flavor.  Three of them $(SU_L(2)$ -- doublets
$\Psi^{(1)}$) interact with the weak boson $W$, the other three
($\Psi^{(2)}$ singlets) are sterile. The doublet states have equal
masses for $u$ and $d$ flavors.

An ordinary, renormalizable, mirror-symmetric theory could form the
basis of the whole construction.

Large masses attributed exclusively to $\Psi_n$ states break the mirror
symmetry. The breakdown leads to parity violation. The $\Psi$ masses
can emerge either dynamically or with asymmetrical Yukawa terms
(interactions with scalar fields). Transition to diagonal forms in the
$\Psi$ mass terms will return Eq.~(19) to a nondiagonal state.
Therefore, the independence of the procedure of an arbitrary choice of
$A,B$ vectors can turn out to be an important property.

Let us touch upon some results that will be generated by any possible
breaking mechanisms and for mirror models of the Lee--Yang type.

The lightest mirror particle has to be much heavier than  the $t$
quark: $M_t\gg m_t$. However mirror generations represent systems
absolutely independent of the usual quark generations. The ratio
$M_t/m_t$ can be much less than those factors of tens and even hundreds
that are exhibited in the ratios of observed quarks. The $\sim1\,$TeV,
or lower region seems to be a good area (see \cite{10} about
perspectives of the LHC observation). Mirror quarks are mainly produced
in pairs, similar to usual quarks. The lightest quark can weakly decay
to yield normal quarks and leptons, but with the coupling constant
$\sim(m/M)^{1/2}$ times less than the weak constant. Production
amplitudes of a unit mirror particle also will be lowered by the same
factor. Such a small value follows from the fact that both usual and
mirror quarks couple in the common current in this very proportion,
i.e. $\sim(m/M)^{1/2}$. The latter factor is connected with the absence
of direct $\bar\psi_{Ra}\psi_{Lb}$ terms in the Lagrangians (a
``see--saw" like situation \cite{11}). Small $R$ contributions
$\sim(m/M)$ can also emerge among the weak currents. Any new phenomena
acquire the largest magnitude in processes with $t$ quark
participation. Neutral currents do not change mirror states to usual
ones.

The quark spectrum formation occurs to be highly similar to the
see--saw-1\,mechanism of the neutrino physics (see review \cite{11}).

I would like to thank Ya.I.~Azimov, M.I.~Eides, G.S.~Danilov, and
L.N.~Lipatov for useful and valuable discussions and suggestions. This
work was supported by grant RSGSS-4801.2012.2.

\end{document}